\documentclass{article}

\PassOptionsToPackage{numbers,sort&compress}{natbib}
\usepackage[final]{neurips_2024_ml4ps}

\usepackage[utf8]{inputenc} 
\usepackage[T1]{fontenc}    
\usepackage{hyperref}       
\usepackage{url}            
\usepackage{booktabs}       
\usepackage{amsfonts}       
\usepackage{nicefrac}       
\usepackage{microtype}      
\usepackage{xcolor}         
\usepackage{amsmath}
\usepackage{graphicx}

\title{Multidimensional Deconvolution with Profiling}

\author{%
  Huanbiao Zhu \\
  Department of Statistics and Data Science\\
  Carnegie Mellon University\\
  Pittsburgh, PA 15213 \\
  \texttt{huanbiaz@andrew.cmu.edu} \\
   \And
   Coauthor \\
}
\author{%
Huanbiao Zhu$^{1}$ \quad Krish Desai$^{2,3}$ \quad Mikael Kuusela$^1$ 
 \\ 
   \quad \textbf{Vinicius Mikuni}$^4$ \quad \textbf{Benjamin Nachman}$^{3,5}$ \quad \textbf{Larry Wasserman}$^{1}$\\
\\
$^1$ Department of Statistics and Data Science, Carnegie Mellon University, Pittsburgh, PA 15213, USA \\
$^2$ Department of Physics, University of California, Berkeley, CA 94720, USA \\
$^3$ Physics Division, Lawrence Berkeley National Laboratory, Berkeley, CA 94720, USA \\
$^4$ National Energy Research Scientific Computing Center, Berkeley Lab, Berkeley, CA 94720, USA \\
$^5$ Berkeley Institute for Data Science, University of California, Berkeley, CA 94720, USA
\\
\texttt{\{huanbiaz,mkuusela\}@andrew.cmu.edu}\\
\texttt{larry@stat.cmu.edu}\\
\texttt{krish.desai@berkeley.edu}\\
\texttt{\{vmikuni,bpnachman\}@lbl.gov}
}

\begin{document}

\maketitle

\begin{abstract}
In many experimental contexts, it is necessary to statistically remove the impact of instrumental effects in order to physically interpret measurements.  This task has been extensively studied in particle physics, where the deconvolution task is called unfolding.  A number of recent methods have shown how to perform high-dimensional, unbinned unfolding using machine learning.  However, one of the assumptions in all of these methods is that the detector response is correctly modeled in the Monte Carlo simulation.  In practice, the detector response depends on a number of nuisance parameters that can be constrained with data.  We propose a new algorithm called Profile OmniFold, which works in a similar iterative manner as the OmniFold algorithm while being able to simultaneously profile the nuisance parameters. We illustrate the method with a Gaussian example as a proof of concept highlighting its promising capabilities. 
\end{abstract}

\section{Introduction}
Instrumental effects distort spectra from their true values.  Statistically removing these distortions is essential for comparing results across experiments and for facilitating broad, detector-independent analysis of the data.  This deconvolution task (called \textit{unfolding} in particle physics) is an ill-posed inverse problem, where small changes in the measured spectrum can result in large fluctuations in the reconstructed true spectrum. In practice, one observes data from the measured spectrum from an experiment, and the goal is to estimate the true spectrum and quantify its uncertainty. See, e.g.,~\cite{Balasubramanian:2019itp, Blobel:2203257,doi:10.1002/9783527653416.ch6, Cowan:2002in, Kuusela2012StatisticalII} for reviews of the problem. The detailed setup will also be introduced in section \ref{setup}.

Traditionally, unfolding has been solved in a discretized setting, where measurements are binned into a histogram (or are naturally represented as discrete, e.g., in images) and the reconstructed spectrum is also represented as a histogram. However, this requires pre-specifying the number of bins, which itself is a tuning parameter and can vary between different experiments. Additionally, binning limits the number of observables that can be simultaneously unfolded.

A number of machine learning-based approaches have been proposed to address this problem~\cite{Arratia:2021otl,Huetsch:2024quz}.  The first one to be deployed to experimental data~\cite{H1:2021wkz,H1prelim-22-031,H1:2023fzk,H1prelim-21-031,LHCb:2022rky,Komiske:2022vxg,Song:2023sxb,Pani:2024mgy,CMS-PAS-SMP-23-008,ATLAS:2024xxl,ATLAS:2024jrp} is OmniFold~(OF)~\cite{Andreassen2020,Andreassen:2021zzk}. Unlike traditional methods, OmniFold does not require binning and can be used to unfold observables in much higher dimensions using neural network (NN) classifiers. The algorithm is an instance of the  Expectation-Maximization (EM) algorithm, which iteratively reweights the simulated events to match the experimental data. The result is guaranteed to converge to the maximum likelihood, provided an infinite sample size and the optimal classifier. However, one limitation in OmniFold, as in all machine learning-based unfolding methods, is the assumption that the detector response is correctly modeled in the simulation\footnote{By 'correctly modeled,' we mean that both the parametric model for the detector response and the nuisance parameters are correctly specified.}. In practice, this is only approximately true, with the simulation depending on a number of nuisance parameters that can be constrained by the observed data.

Recently, \cite{Chan2023} introduced an unbinned unfolding method that also allows for profiling the nuisance parameters. This is achieved by using machine learning to directly maximize the log-likelihood function.  While a significant step forward, this approach is limited to the case where the detector-level data are binned so that one can write down the explicit likelihood (each bin is Poisson distributed).

In this work, we propose a new algorithm, called Profile OmniFold (POF), for unbinned and profiled unfolding. Unlike \cite{Chan2023}, POF is completely unbinned at both the detector-level and pre-detector-level (`particle level').  Additionally, POF can be seen as an extension to the original OF algorithm that iteratively reweights the simulated particle-level events but also simultaneously determines the nuisance parameters.

\section{Methodology}
\label{method}
In this section, we introduce POF, which is a modified version of the original OF algorithm. Same as the original OF, the goal of POF is to find the maximum likelihood estimate of the reweighting function that reweights generated particle-level data from $q(x)$ to the truth $p(x)$. However, unlike in the original OF algorithm, POF can also take into account nuisance parameters in the detector modeling and simultaneously profile out these nuisance parameters. At the same time, POF retains the same benefits as OF such that it can directly work with unbinned data, utilize the power of NN classifiers, and unfold multidimensional observables or even the entire phase space simultaneously~\cite{Andreassen2020}.

\subsection{Statistical setup of the unfolding problem in the presence of nuisance parameters}
\label{setup}
In the unfolding problem, we are provided pairs of Monte Carlo (MC) simulations $\{X_i,Y_i\}_{i=1}^n\sim q(x,y)$, where $X_i$ denotes a particle-level quantity and $Y_i$ the corresponding detector-level observation. Then given a set of observed detector-level experimental data $\{Y_i'\}_{i=1}^m \sim p(y)$, our goal is to estimate the true particle-level density $p(x)$. The forward model for both MC simulation and experimental data are described by
\begin{equation}
\label{eq:forward}
    q(y) = \int q(y|x)q(x)dx, \;\; p(y) = \int p(y|x)p(x)dx,
\end{equation}
where $q(y|x)$ and $p(y|x)$ are kernels that represent the detector responses. In practice, different detector configurations yield different detector responses, so it is often the case that $q(y|x)\neq p(y|x)$. Additionally, the response kernel is assumed to be parameterized by some nuisance parameters $\theta$, which are given for the MC data but unknown for the experimental data.

Given this setup, let $\nu(x)$ be a reweighting function on the MC particle-level density $q(x)$. Ultimately, we want $\nu(x)\approx p(x)/q(x)$. Also, suppose $q(y|x)$ is specified by nuisance parameter $\theta_0$, i.e. $q(y|x)=p(y|x,\theta_0)$. Let $w(y,x,\theta)$ be a reweighting function on the MC response kernel $q(y|x)$, i.e., $w(y,x,\theta)=p(y|x,\theta)/q(y|x)$ Then the goal is to maximize the population log-likelihood
\begin{equation}
\label{eq:log_likelihood}
\begin{split}
    \ell(\nu,\theta) &= \int p(y)\log p(y|\nu,\theta)dy + \log p_0(\theta) \\
    &\text{subject to }\int \nu(x)q(x)dx=1,
\end{split}
\end{equation}
where $p_0(\theta)$ is a prior on $\theta$ to constrain the nuisance parameter, usually determined from auxiliary measurements. In our case, we use the standardized Gaussian prior, $\log p_0(\theta)=-\frac{\|\theta-\theta_0\|^2}{2}$. 

\subsection{Algorithm} 
\label{algorithm}
The POF algorithm, like the original OF algorithm, is an EM algorithm. It iteratively updates the reweighting function $\nu(x)$ and nuisance parameter $\theta$ toward the maximum likelihood estimate. The key in the EM algorithm is the $Q$ function, which is the complete data ($x,y$) expected log-likelihood given the observed data ($y$) and current parameter estimates ($\nu^{(k)},\theta^{(k)}$). For the log-likelihood specified in \eqref{eq:log_likelihood}, the $Q$ function is given by
\begin{equation}
\begin{split}
    Q(\nu,\theta|\nu^{(k)},\theta^{(k)}) &= \int p(y)\int p(x|y,\nu^{(k)},\theta^{(k)})\log p(y,x|\nu,\theta)dxdy + \log p_0(\theta) \\
    &\text{subject to }\int \nu(x)q(x)dx=1.
\end{split}
\end{equation}
The E-step in the EM algorithm is to compute the $Q$ function and M-step is to maximize over $\nu$ and $\theta$. The maximizer will then be used as the updated parameter values in the next iteration. Specifically, in the $k^{\text{th}}$ iteration, we obtain the update $(\nu^{(k+1)},\theta^{(k+1)})$ by solving $(\nu^{(k+1)},\theta^{(k+1)})={\arg\max}_{\nu,\theta}Q(\nu,\theta|\nu^{(k)},\theta^{(k)})$. As shown in Appendix~\ref{app:derivation}, we can solve this optimization problem in three steps:
\begin{enumerate}
    \item $r^{(k)}(y) = \frac{p(y)}{\tilde{q}^{(k)}(y)}$, \\
    \text{where }$\tilde{q}^{(k)}(y)=\int w(y,x,\theta^{(k)})\nu^{(k)}(x)q(y,x)dx$
    \item $\nu^{(k+1)}(x) = \nu^{(k)}(x)\frac{\tilde{q}^{(k)}(x)}{q(x)}$, \\
    \text{where } $\tilde{q}^{(k)}(x)=\int w(y,x,\theta^{(k)})r^{(k)}(y)q(y,x)dy$ \\
    \item Find $\theta^{(k+1)}$ such that $\theta^{(k+1)}-\theta_0 =  \int\int w(y,x,\theta^{(k)})\nu^{(k)}(x)\frac{\Dot{w}(y,x,\theta^{(k+1)})} {w(y,x,\theta^{(k+1)})}r^{(k)}(y)q(y,x)dxdy$
\end{enumerate}
The first step is almost the same as the first step in the original OF algorithm, which involves computing the ratio of the detector-level experimental density and reweighted detector-level MC density using the push-forward weights of $w(y,x,\theta^{(k)})\nu^{(k)}(x)$. The density ratio can be estimated by training a NN classifier to distinguish between the experimental data distribution $p(y)$ and the reweighted MC distribution $\tilde{q}^{(k)}(y)$. 

The second step also closely mirrors the second step of the original OF algorithm and involves computing the ratio of the reweighted particle-level MC density using the pull-back weights of $w(y,x,\theta^{(k)})r^{(k)}(y)$ and the particle-level MC density.

The third step involves updating the nuisance parameter numerically. The right-hand side of the equation is more involved, since it requires computing $\frac{\Dot{w}(y,x,\theta)} {w(y,x,\theta)}$, where $\Dot{w}(y,x,\theta)$ is the derivative of $w(y,x,\theta)$ with respect to $\theta$. Fortunately, \cite{Chan2023} shows that the dependency of $w(y,x,\theta)$ on $\theta$ can be learned through neural conditional reweighting~\cite{Nachman2022} using another set of synthetic data $(X_i,Y_i,\theta_i)$. Then, the trained network provides estimates for both $w(y,x,\theta)$ and its derivative $\Dot{w}(y,x,\theta)$. Additionally, $w(y,x,\theta^{(k)}),\nu^{(k)}(x),r^{(k)}(y)$ have all been computed in the previous steps. Finally, the integral is over the joint distribution $q(y,x)$ so we can just use the empirical average to obtain the estimate. 

In summary, the POF algorithm extends the original OF iteration by including an additional step for updating the nuisance parameter. However, unlike OF, POF does not have guaranteed convergence to the population maximum likelihood since the likelihood function might not be concave in $\theta$. The algorithm iterates through these three steps for a finite number of iterations, typically fewer than 10. Early stopping is often used to help regularize the solution.

\section{Gaussian example}
\label{Gaussian}
We illustrate the POF algorithm with a simple Gaussian example.  Consider a one-dimensional Gaussian distribution at the particle level and two Gaussian distributions at the detector level. The data are generated as follows:
\begin{align*}
    Y_{i1} &= X_{i} + Z_{i1}, \\
    Y_{i2} &= X_{i} + Z_{i2},
\end{align*}
where $X_i\sim\mathcal{N}(\mu,\sigma^2), Z_{i1}\sim\mathcal{N}(0,1), Z_{i2}\sim\mathcal{N}(0,\theta^2)$. Here, $\theta$ is the nuisance parameter, which only affects the second dimension of the detector-level data.  This is qualitatively similar to the physical case of being able to measure the same quantity twice.  Since the response kernel in this case is a Gaussian density, we have access to the analytic form $p(y|x,\theta)$ and, consequently, $w(y,x,\theta)$ as well. For simplicity, we use the analytic form directly in the algorithm for this example. However, even if we do not know the analytic form, we can learn $w(y,x,\theta)$ as described in Sec.~\ref{algorithm}.

\paragraph{Dataset} Based on the above data-generating process, Monte Carlo data are generated with $\mu=0,\sigma=1,\theta=1$ and experimental data are generated with $\mu=0.8,\sigma=1,\theta=1.5$. We simulate $10^5$ events each for the MC data and experimental data.

\paragraph{Neural network architecture and training}
The neural network classifier for estimating the density ratios is implemented using TensorFlow and Keras. The network contains three hidden layers with 50 nodes per layer and employs the ReLU activation function. The output layer consists of a single node with a sigmoid activation function. The network is trained using the Adam optimizer~\cite{kingma2017adammethodstochasticoptimization} to minimize the weighted binary cross-entropy loss. The network is trained for 10 epochs with a batch size of 10000.  None of these parameters were optimized for this proof-of-concept demonstration. All training was performed on an NVIDIA A100 Graphics Processing Unit (GPU), taking no more than 10 minutes.

\begin{figure}[h]
  \centering
  \includegraphics[width=0.55\linewidth]{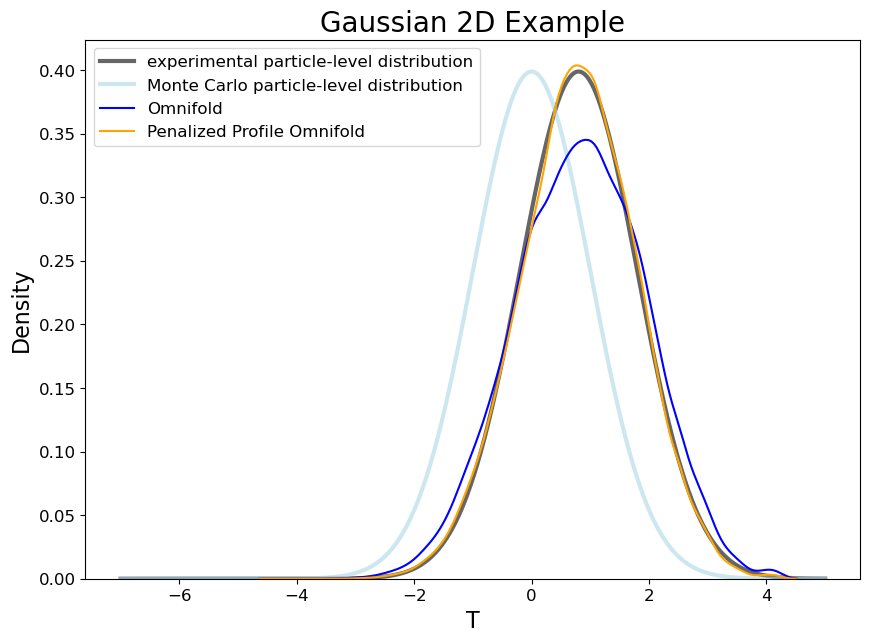}
  \includegraphics[width=0.4\linewidth]{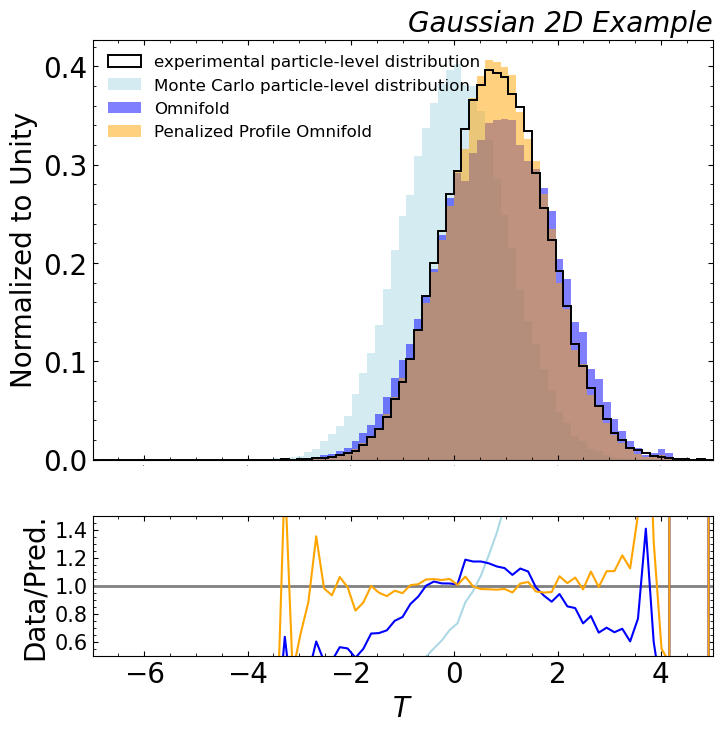}
  \caption{Results of unfolding a 2D Gaussian example. Left: The unfolded particle-level density using POF ({\color{orange}orange}) and OF ({\color{blue}blue}), with both algorithms running for 5 iterations. Top-right: Unfolded spectrum aggregated into 80 bins. Bottom-right: Ratio of the unfolded spectrum to the truth spectrum.} 
  \label{fig:Gaussian2DExample}
\end{figure}

\subsection{Results}

Figure \ref{fig:Gaussian2DExample} illustrates the results of unfolding the 2D Gaussian data using both the proposed POF algorithm and the original OF algorithm. The cyan line is the Monte Carlo distribution for which the reweighting function $\nu(x)$ is applied. The results show that the original OF algorithm (blue line) deviates significantly from the true distribution (black line). This discrepancy arises because OF assumes $p(y|x)=q(y|x)$, an assumption that is invalid in the presence of incorrectly specified nuisance parameters. On the other hand, the POF algorithm simultaneously optimizes the nuisance parameter along with the reweighting function. The results show that the unfolded solution (orange line) aligns closely with the truth (black line) and the estimated nuisance parameter is $\hat{\theta}=1.48$ (true parameter is $\theta = 1.50$).  Future work will deploy standard techniques like bootstrapping to determine uncertainties.

\section{Conclusion}
\label{conclusion}
In this work, we have proposed a new algorithm called Profile OmniFold, which uses machine learning to perform unfolding while also simultaneously profiling out the nuisance parameters. This relaxes the original assumption in OmniFold that the detector response needs to be correctly modeled in the Monte Carlo simulation and constrains the nuisance parameter in a data-driven way. At the same time, the proposed algorithm still shares similar steps as in OF preserving its many benefits, including ease of implementation. 

The results from the simple Gaussian example demonstrate the algorithm's promising performance. Our next objective is to apply POF to more realistic examples and include critical studies on robustness, stability, and uncertainty quantification.

\begin{ack}
We thank Jesse Thaler for many useful discussions about OmniFold and related subjects as well as feedback on the manuscript.  KD, VM, BN, and HZ were supported by the U.S. Department of Energy (DOE), Office of Science under contract DE-AC02-05CH11231. HZ, MK and LW were partially supported by National Science Foundation grants DMS-2310632 and DMS-2053804. This research used resources of the National Energy Research Scientific Computing Center, a DOE Office of Science User Facility supported by the Office of Science of the U.S. Department of Energy under Contract No. DE-AC02-05CH11231 using NERSC award HEP-ERCAP0021099. 

\end{ack}

\bibliography{myrefs}


\appendix

\section{Appendix / supplemental material} \label{app:derivation}

In this appendix, we provide the derivation of the POF algorithm presented in Sec.~\ref{algorithm}. As mentioned there, in each iteration, the algorithm aims to maximize the Q-function
\begin{align*}
    Q(\nu,\theta|\nu^{(k)},\theta^{(k)}) &= \int p(y)\int p(x|y,\nu^{(k)},\theta^{(k)})\log p(y,x|\nu,\theta)dxdy + \log p_0(\theta) \\
    &=\int p(y)\int p(x|y,\nu^{(k)},\theta^{(k)})\log[w(y,x,\theta)q(y|x)\nu(x)q(x)]dxdy + \log p_0(\theta) \\
    &\text{subject to }\int \nu(x)q(x)dx=1.
\end{align*}
For simplicity, consider the case where $\theta\in\mathbb{R}$, but the argument can also be extended to higher dimensions. Assume the prior is a standard Gaussian, $\log p_0(\theta)=-\frac{(\theta-\theta_0)^2}{2}$, and write the Q-function in its Lagragian form,
\begin{align*}
    \tilde{Q}(\nu,\theta|\nu^{(k)},\theta^{(k)}) &= Q(\nu,\theta|\nu^{(k)},\theta^{(k)}) - \lambda_1\left(\int \nu(x)q(x)dx-1\right). 
\end{align*}
First, we take derivative of $\tilde{Q}$ with respect to $\nu(x)$ and set it to be 0,
\begin{align*}
    \frac{\delta}{\delta\nu(x)}\tilde{Q}(\nu,\theta|\nu^{(k)},\theta^{(k)})=\frac{\int p(y)p(x|y,\nu^{(k)},\theta^{(k)})dy}{\nu(x)} - \lambda_1 q(x) = 0.
\end{align*}
Integrating both sides over $\int \nu(x)dx$ yields that $\lambda_1=1$.
Therefore, the stationary condition for $\nu(x)$ satisfies
\begin{equation}
\label{eq:nu_stationary}
\begin{split}
    \nu(x) &= \frac{\int p(y)p(x|y,\nu^{(k)},\theta^{(k)})dy}{q(x)} \\
    &= \int \frac{p(y)w(y,x,\theta^{(k)})q(y|x)\nu^{(k)}(x)dy}{\int w(y,x',\theta^{(k)})q(y|x')\nu^{(k)}(x')q(x')dx'} \\
    &= \nu^{(k)}(x)\int w(y,x,\theta^{(k)})q(y|x)\frac{p(y)}{\tilde{q}^{(k)}(y)}dy,
\end{split}
\end{equation}
where $\tilde{q}^{(k)}(y)=\int w(y,x',\theta^{(k)})q(y|x')\nu^{(k)}(x')q(x')dx'$. Multiplying the right-hand side by $\frac{q(x)}{q(x)}$, we obtain
\begin{align*}
    \nu(x) &= \nu^{(k)}(x)\frac{\int w(y,x,\theta^{(k)})\frac{p(y)}{\tilde{q}^{(k)}(y)}q(y,x)dy}{q(x)}.
\end{align*}
This corresponds to exactly the first two steps in the algorithm, where $r^{(k)}(y)=\frac{p(y)}{\tilde{q}^{(k)}(y)}$ and $\tilde{q}^{(k)}(x)=\int w(y,x,\theta^{(k)})r^{(k)}(y)q(y,x)dy$. On the other hand, taking derivative of $\tilde{Q}$ with respect to $\theta$, we obtain
\begin{align*}
    \frac{\delta}{\delta \theta}\tilde{Q}(\nu,\theta|\nu^{(k)},\theta^{(k)})
    &= \int\int\frac{p(y)p(x|y,\nu^{(k)},\theta^{(k)})\Dot{w}(y,x,\theta)} {w(y,x,\theta)}dxdy - (\theta-\theta_0)
\end{align*}
where $\Dot{w}(y,x,\theta)=\frac{\delta w(y,x,\theta)}{\delta\theta}$. Setting this to be 0, we have
\begin{equation}
\label{eq:theta_stationary}
\begin{split}
    \theta-\theta_0 &=  \int\int\frac{p(y)p(x|y,\nu^{(k)},\theta^{(k)})\Dot{w}(y,x,\theta)} {w(y,x,\theta)}dxdy  \\
    &= \int\int\frac{p(y)w(y,x,\theta^{(k)})q(y|x)\nu^{(k)}(x)q(x)\Dot{w}(y,x,\theta)} {w(y,x,\theta)\int w(y,x',\theta^{(k)})q(y|x')\nu^{(k)}(x')q(x')dx'}dxdy \\
    &= \int\int q(y,x)w(y,x,\theta^{(k)})\nu^{(k)}(x)\frac{\Dot{w}(y,x,\theta)} {w(y,x,\theta)}\frac{p(y)}{\tilde{q}^{(k)}(y)}dxdy,
    \end{split}
\end{equation}
where again $\tilde{q}^{(k)}(y)=\int w(y,x',\theta^{(k)})q(y|x')\nu^{(k)}(x')q(x')dx'$. This corresponds to the third step in the algorithm with $r^{(k)}(y)=\frac{p(y)}{\tilde{q}^{(k)}(y)}$. While this step requires the assumption that $\tilde{Q}(\nu,\theta|\nu^{(k)},\theta^{(k)})$ is concave in $\theta$, which is not always guaranteed, one can also consider alternative ways for updating $\theta$. For example, instead of solving for the stationary point of $\tilde{Q}$ with respect to $\theta$, one could employ a first-order update
\begin{align*}
    \theta^{(k+1)} &= \theta^{(k)} + \gamma\cdot\frac{\delta}{\delta \theta}\tilde{Q}(\nu,\theta|\nu^{(k)},\theta^{(k)})|_{\theta=\theta^{(k)}},
\end{align*}
where $\gamma\geq 0$ is an appropriately chosen step size. This approach effectively performs a single step of gradient ascent with respect to $\theta$.

\end{document}